\documentclass[unsortedaddress,preprint]{revtex4-1}

\usepackage{amsmath,amssymb,color,graphicx}

\begin{document}

\title{
Azimuthal angle distributions of neutrons emitted from the $^9$Be($\gamma$,$n$) reaction with linearly polarized $\gamma$-rays
}



\author{Yuma Kikuchi}
\email[]{yuma@sci.osaka-cu.ac.jp}
\affiliation{Department of Physics, Graduate School of Science, Osaka City University, Osaka 558-8585, Japan}

\author{Kazuyuki Ogata}
\affiliation{Research Center for Nuclear Physics (RCNP), Osaka University, Ibaraki 567-0047, Japan}
\affiliation{Department of Physics, Graduate School of Science, Osaka City University, Osaka 558-8585, Japan}

\author{Takehito Hayakawa}
\affiliation{National Institutes for Quantum and Radiological Science and Technology, 2-4 Shirakata, Tokai, Naka, Ibaraki 319-1106, Japan}

\author{Satoshi Chiba}
\affiliation{Tokyo Institute of Technology, 2-12-1 Ookayama, Meguro, Tokyo 113-0033, Japan}

\date{\today}

\begin{abstract}
The electromagnetic transitions of $^9$Be with linearly polarized $\gamma$-rays are calculated by using the $\alpha$~+~$\alpha$~+~$n$ three-body model and the complex-scaled solutions of the Lippmann-Schwinger equation; the azimuthal angle distributions of the emitted neutrons are investigated.
We calculate the anisotropy parameter as a function of the photon incident energy $E_\gamma$, and discuss how sensitive the anisotropy parameter is to nuclear structure and transition modes.
The result suggests that the azimuthal angle distribution of neutrons emitted from the $^9$Be($\gamma$,$n$) reaction with the linearly polarized $\gamma$-rays is useful to identify the resonances in the final states even if it is not clearly observed in the cross section.
\end{abstract}



\maketitle

\section{Introduction}

The photonuclear reactions have an important role in developing nuclear physics~\cite{Heyde10}.
In particular, the ($\gamma$, $n$) reactions with linearly polarized $\gamma$-ray beams have the potential to allow us to study the nuclear structures in detail.
In 1957, Agodi~\cite{Agodi57} predicted that azimuthal angle distributions of nucleons emitted from the ($\vec{\gamma}$, $n$) and ($\vec{\gamma}$, $p$) reactions have anisotropic shapes and are proportional to the function of $1+b\cos{(2\phi)}$ at the polar angle $\theta = 90^\circ$, where $\phi$ is the azimuthal angle.
The coefficient $b$ contains the information on the transition modes and nuclear structure of excited states such as the single-particle structure of emitted nucleons.

Recent development in experimental technique of laser Compton scattering (LCS) enables us to investigate the ($\vec{\gamma}$,$n$) and ($\vec{\gamma}$,$p$) reactions.
An advantage of LCS is that one can generate almost 100~\% linearly polarized $\gamma$-ray beams because the polarization of the laser is directly transferred to the photons.
The linearly polarized $\gamma$-ray generated by LCS is now available at HI$\gamma$S~\cite{Arnold12,Mueller15}, NewSUBARU~\cite{Horikawa14,Utsunomiya15,Hayakawa16}, and so on, and in fact, the azimuthal angle distributions of neutrons emitted from the ($\vec{\gamma}$,$n$) reactions were measured by several groups~\cite{Mueller15,Horikawa14,Hayakawa16}.

Theoretically, there is no detailed discussion on the ($\vec{\gamma}$,$n$) and ($\vec{\gamma}$,$p$) reactions in relation with nuclear structure.
In Ref.~\cite{Agodi57}, the azimuthal angle distribution was discussed for the electric and magnetic transitions, but the nuclear structure dependence, which can be described as the single-particle structure of emitted nucleons, was not taken into account in the anisotropic shapes in the distributions.
To compare with experimental data, one should investigate the azimuthal angle distribution of emitted nucleons in relation with the nuclear structure and transition modes.

The $^9$Be($\vec{\gamma}$,$n$) reaction is one of the best examples for such studies because the reaction reveals different aspects of structure of $^9$Be in continuum, depending on the excitation energy.
In a low-energy region, the cross section of the $^9$Be($\gamma$,$n$) reaction has peaks coming from the resonances connected with the ground state by the $E1$ and $M1$ transitions~\cite{Garrido11,Casal14,Odsuren15,Kikuchi16a}.
Thus, the azimuthal angle distributions in the low energy region tell us the information for each resonance. 
In a higher energy region below the giant dipole resonance, the reaction is dominated by the transitions into non-resonant continuum states~\cite{Kikuchi16a}, which are expected to have no peculiar structure in the final states.
From the azimuthal angle distributions in the higher energy region, we can discuss the sensitivity of the anisotropy to the transition modes.

The purpose of this work is to investigate the azimuthal angle distribution of the neutrons emitted from the $^9$Be($\vec{\gamma}$,$n$) reaction and to present the anisotropy in the distribution in relation with the nuclear structure and transition modes.
To calculate the $^9$Be($\vec{\gamma}$,$n$) reaction, we use the $\alpha$~+~$\alpha$~+~$n$ three-body model~\cite{Odsuren15,Kikuchi16a}, which reproduces the cross section of the $^9$Be($\gamma$,$n$) reaction quantitatively.
For final states, it is required to describe the three-body scattering states of the $\alpha$~+~$\alpha$~+~$n$ system.
We here describe the final scattering states by combining the three-body model with the complex-scaled solutions of the Lippmann-Schwinger equation (CSLS)~\cite{Kikuchi09}.
To discuss the nuclear structure and transition modes from the azimuthal angle distributions, we calculate the coefficient $b$ as a function of the photon incident energy.
We show the coefficients $b$ at energies corresponding to the resonances, which are excited by the $E1$ and $M1$ transitions, and discuss whether or not the azimuthal angle distributions reflect the nuclear structure. 
We also show the anisotropy for the transitions into the non-resonant continuum states, and discuss how sensitive the anisotropy is to the transition modes.

\section{Theoretical framework}
\subsection{$\alpha$~+~$\alpha$~+~$n$ three-body model}

To describe the $^9$Be system, we employ the $\alpha$~+~$\alpha$~+~$n$ three-body orthogonality condition model (OCM)~\cite{Kikuchi16a}.
In the OCM, the Hamiltonian for relative motions of the $\alpha$~+~$\alpha$~+~$n$ system is given by
\begin{equation}
H = \sum_{i=1}^3 t_i - T_\text{cm} + \sum_{i=1}^2 V_{\alpha n} (\boldsymbol{\xi}_i) + V_{\alpha\alpha} + V_\text{PF} + V_{\alpha\alpha n},
\label{eq:hami}
\end{equation}
where $t_i$ and $T_\text{cm}$ are kinetic energies for individual particles and the center-of-mass of the system, respectively.
The interaction between the neutron and the $i$-th $\alpha$ particle is given by $V_{\alpha n}(\boldsymbol{\xi}_i)$, where $\boldsymbol{\xi}_i$ is the relative coordinate between them.
Here, we employ the KKNN potential~\cite{Kanada79} for $V_{\alpha n}$.
For the $\alpha$-$\alpha$ interaction $V_{\alpha\alpha}$, we use the same potential as used in Ref.~\cite{Kurokawa05}, which is a folding potential of the effective $NN$ interaction~\cite{Schmid61} and the Coulomb interaction.
The explicit form of $V_{\alpha\alpha}$ is given by
\begin{equation}
V_{\alpha\alpha} = V_N \exp{(-\mu_{\alpha\alpha} r^2)} + \frac{4e^2}{r} \text{erf}\left( -\kappa r \right),
\label{eq:vaa}
\end{equation}
whose parameters are given in Ref.~\cite{Odsuren15}.
The pseudo potential $V_\text{PF}$~\cite{Kukulin86} is in fact the projection operator
\begin{equation}
V_\text{PF} = \lambda | \Phi_\text{PF} \rangle \langle \Phi_\text{PF} |,
\end{equation}
which removes the Pauli forbidden states from the relative motions of $\alpha$-$\alpha$ and $\alpha$-$n$ subsystems.
The Pauli forbidden states are defined by the harmonic oscillator wave functions by assuming the $(0s)^4$ configuration for the $\alpha$ particle.
In the present calculation, we take $\lambda$ as $10^6$ MeV.
In the present model, we introduce the phenomenological $\alpha$~+~$\alpha$~+~$n$ three-body potential $V_{\alpha\alpha n}$~\cite{Kikuchi16a}.
The explicit form is given by
\begin{equation}
V_{\alpha\alpha n} = V_3 \exp{\left( -\mu_3 \rho^2 \right)},
\end{equation}
where $\rho$ is the hyperradius of the $\alpha$~+~$\alpha$~+~$n$ system.
The strength and width parameters of the three-body potential, $V_3$ and $\mu_3$, are determined for each spin-parity state.
For $3/2^-$ states, we determine the parameters to reproduce the observed binding energy and charge radius of the ground state because these quantities are important to reproduce the $Q$-value and sum rule values of the electric dipole transition; we take $V_3 = 1.10$ MeV and $\mu = 0.02$ fm$^{-2}$.
For other spin-parity states, we use the same value of $\mu$ as used for the $3/2^-$ states, whereas the $V_3$ are so as to reproduce the peak energies of the photodisintegration cross section.

With the Hamiltonian in Eq.~(\ref{eq:hami}), we consider the following Schr\"odinger equation
\begin{equation}
H \chi^{J^\pi}_\nu = E_\nu \chi^{J^\pi}_\nu,
\label{eq:sceq}
\end{equation}
where $J^\pi$ is the total spin and the parity of the $\alpha$~+~$\alpha$~+~$n$ system.
The energy eigenvalue and the eigenstate of the relative motions of the system are expressed by $E_\nu$ and $\chi^{J^\pi}_\nu$, respectively, in which $\nu$ is the state index.
To solve the Schr\"odinger equation, we employ the coupled-rearrangement-channel Gaussian expansion method~\cite{Hiyama03}.
In the present calculation, we describe the relative wave function $\chi^{J^\pi}_\nu$ as
\begin{equation}
\chi^{J^\pi}_\nu = \sum_{i j c} \mathcal{C}_{ijc}^\nu(J^\pi) \left[\left[\phi_l^i (\mathbf{r}_c) \otimes \phi_\lambda^j (\mathbf{R}_c)\right]_L\otimes \chi^\sigma \right]_{JM},
\end{equation}
where $\mathcal{C}_{ijc}^\nu(J^\pi)$ is the expansion coefficient and $\chi^\sigma$ is the spin wave function of the neutron.
The relative coordinates $\mathbf{r}_c$ and $\mathbf{R}_c$ are those in three kinds of Jacobi coordinate systems labeled by $c$ ($c=1,2,3$), and the indices for the basis functions are represented by $i$ and $j$.
The spatial part of the wave functions is expanded with Gaussian basis functions given by
\begin{equation}
\phi_l^i (\mathbf{r}) = N_l^i r^l \exp{\left(-\frac{1}{2}a_i r^2 \right)}Y_l (\hat{\mathbf{r}}),
\end{equation}
where $N_l^i$ is a normalization factor and $a_i$ is the width of the Gaussian.

\subsection{Complex-scaled solutions of the Lippmann-Schwinger equation}
To investigate the $^9$Be($\gamma$,$n$) reaction, it is necessary to describe the three-body scattering states of $\alpha$~+~$\alpha$~+~$n$.
We adopt the complex-scaled solutions of the Lippmann-Schwinger equation (CSLS)~\cite{Kikuchi09}, in which the complex scaling method (CSM) is combined with the Lippmann-Schwinger formalism.
Before going into the formalism of CSLS, we briefly explain CSM~\cite{Aguilar71,Balslev71,Ho83,Moiseyev98,Aoyama06,Myo14}.
In CSM, the relative coordinates $\boldsymbol{\zeta} = (\mathbf{r}_c, \mathbf{R}_c)$ are transformed as
\begin{equation}
U(\theta) \boldsymbol{\zeta} U^{-1}(\theta) = \boldsymbol{\zeta} e^{i\theta},
\end{equation}
where $U(\theta)$ is the complex scaling operator with a scaling angle $\theta$ being a real number.
Applying this transformation to the Hamiltonian $H$, we obtain the complex-scaled Schr\"odinger equation
\begin{equation}
H^\theta \chi^\theta_\nu = E^\theta_\nu \chi^\theta_\nu,
\label{eq:cssc}
\end{equation}
where $H^\theta$ is the complex-scaled Hamiltonian.
By solving the complex-scaled Schr\"odinger equation with a finite number of $L^2$ basis functions such as Gaussian, we obtain the eigenstates $\{\chi^\theta_\nu\}$ and the energy eigenvalues $\{E^\theta_\nu\}$ of $H^\theta$.

All the energy eigenvalues $\{E^\theta_\nu\}$ are obtained on a complex energy plane, governed by the ABC theorem~\cite{Aguilar71,Balslev71}, and the distributions of their imaginary parts reflect the outgoing boundary conditions as follows.
In CSM, the resonances of a many-body system are obtained as the isolated poles with the $L^2$ basis functions.
The energy eigenvalues of the resonances are given by $E = E_r - i\Gamma/2$, where $E_r$ and $\Gamma$ are the resonance energy and the decay width, respectively.
In contrast, the energy eigenvalues of continuum states are obtained on the $2\theta$-rotated branch cuts starting from different thresholds of two and three-body decay channels, such as $^8$Be~+~$n$ and $\alpha$~+~$\alpha$~+~$n$ in the case of $^9$Be.
This classification of continuum states in CSM imposes that the outgoing boundary condition for each open channel is taken into account automatically by the imaginary parts of energy eigenvalues.
Using the classification of continuum states in CSM, we can describe three-body scattering states without any explicit enforcement of boundary conditions.

The complex-scaled eigenstates satisfy the extended completeness relation~\cite{Myo98}, consisting of bound states, resonances, and rotated continua, as
\begin{equation}
1 = \int\hspace{-0.48cm}\sum_{\nu} | \chi^\theta_\nu \rangle \langle \tilde{\chi}^\theta_\nu |,
\label{eq:ecr}
\end{equation}
where $\{\chi^\theta_\nu, \tilde{\chi}^\theta_\nu\}$ form a set of biorthogonal states.
This relation is used when we describe the scattering states in CSLS.

In CSLS, we start with the formal solution of the Lippmann-Schwinger equation given by
\begin{equation}
\Psi^{(\pm)} (\mathbf{k},\mathbf{K}) = \Phi_0 (\mathbf{k},\mathbf{K}) + \lim_{\varepsilon \to 0} \frac{1}{E-H\pm i\varepsilon} V \Phi_0 (\mathbf{k},\mathbf{K}),
\label{eq:LSfs}
\end{equation}
where $\mathbf{k}$ is the relative momentum between two $\alpha$'s and $\mathbf{K}$ is that between the neutron and the center-of-mass of the $\alpha$-$\alpha$ subsystem.
The function $\Phi_0$ is a solution of the asymptotic Hamiltonian $H_0$ for the $\alpha$~+~$\alpha$~+~$n$ three-body system.
The interaction $V$ in the second term in Eq.~(\ref{eq:LSfs}) is defined by subtracting $H_0$ from $H$.

In the present calculation of the scattering states, for simplicity, we replace the Coulomb part of the $\alpha$-$\alpha$ interaction in Eq.~(\ref{eq:vaa}) with the shielded Coulomb potential with the Gaussian damping factor given by
\begin{equation}
\frac{4e^2}{r}\text{erf}\left(-\kappa r\right) \exp{\left(-\frac{r^2}{R_c^2}\right)}.
\label{eq:shco}
\end{equation}
The parameter $R_c$ is taken as $R_c^2 = 10^7$ fm$^2$.
We have confirmed that the photodisintegration cross section calculated with the shielded Coulomb potential is identical to the original result in Ref.~\cite{Kikuchi16a}.
Then $H_0$ is defined by the kinetic energy operator and its solution is given by
\begin{equation}
\langle \mathbf{r},\mathbf{R} | \Phi_0 (\mathbf{k}, \mathbf{K}) \rangle = \frac{1}{(2\pi)^3} e^{i\mathbf{k}\cdot\mathbf{r} + i\mathbf{K}\cdot\mathbf{R}},
\end{equation}
where $\mathbf{r}$ and $\mathbf{R}$ are the relative coordinates being conjugate to $\mathbf{k}$ and $\mathbf{K}$, respectively.

To describe the electromagnetic transition into the $\alpha$~+~$\alpha$~+~$n$ three-body scattering states, we consider the incoming scattering states in the bra-representation.
Assuming the hermiticities of $H$ and $V$, the scattering states are written as
\begin{equation}
\langle \Psi^{(-)}(\mathbf{k},\mathbf{K}) | = \langle \Phi_0(\mathbf{k},\mathbf{K}) |
+ \lim_{\varepsilon \to 0} \langle \Phi_0(\mathbf{k},\mathbf{K})| V \frac{1}{E-H+i\varepsilon}.
\label{eq:LSbr}
\end{equation}
In CSLS, we express Green's function in Eq.~(\ref{eq:LSbr}) in terms of the complex-scaled Green's function.
The complex-scaled Green's function with the outgoing boundary condition, $\mathcal{G}^\theta(E)$, is related with the non-scaled Green's function as
\begin{equation}
\lim_{\varepsilon \to 0}\frac{1}{E-H+i\varepsilon} = U^{-1}(\theta) \mathcal{G}^\theta(E) U(\theta).
\label{eq:GFre}
\end{equation}
The explicit form of $\mathcal{G}^\theta(E)$ is defined by
\begin{equation}
\mathcal{G}^\theta(E) = \frac{1}{E-H^\theta} = \int\hspace{-0.48cm}\sum_\nu \frac{|\chi^\theta_\nu\rangle \langle \tilde{\chi}^\theta_\nu |}{E-E^\theta_\nu},
\label{eq:csGF}
\end{equation}
where the completeness relation in Eq.~(\ref{eq:ecr}) is used.
From Eqs.~(\ref{eq:LSbr}), (\ref{eq:GFre}), and (\ref{eq:csGF}), we obtain the incoming scattering states in CSLS as
\begin{equation}
\begin{split}
\langle \Psi^{(-)}&(\mathbf{k},\mathbf{K}) | = \langle \Phi_0(\mathbf{k},\mathbf{K}) | \\
&+\int \hspace{-0.48cm}\sum_\nu \langle \Phi_0(\mathbf{k},\mathbf{K}) | V U^{-1}(\theta) | \chi^\theta_\nu \rangle \frac{1}{E-E^\theta_\nu} \langle \tilde{\chi}^\theta_\nu | U(\theta).
\end{split}
\end{equation}

\subsection{Electromagnetic transitions with linearly polarized $\gamma$-rays}

To calculate the photodisintegration cross section with the $\gamma$-ray linearly polarized to the $x$-axis, we consider the following matrix elements:
\begin{equation}
\mathcal{M}_x(EM1) = \left\langle \Psi^{(-)}(\mathbf{k},\mathbf{K}) \Big| \hat{O}_x(EM1) \Big| \Phi_\text{g.s.} \right\rangle,
\label{eq:mele}
\end{equation}
where $\Phi_\text{g.s.}$ is the initial ground-state wave function and $\hat{O}_x(EM1)$ is the electromagnetic dipole transition operator.
The operators for the electric and magnetic transitions are defined by
\begin{equation}
\hat{O}_x(E1) = -\frac{1}{\sqrt{2}}\left(\hat{O}_{11}(\mathcal{E}) - \hat{O}_{1-1}(\mathcal{E})\right)
\end{equation}
and
\begin{equation}
\hat{O}_x(M1) = \frac{i}{\sqrt{2}}\left(\hat{O}_{11}(\mathcal{M}) + \hat{O}_{1-1}(\mathcal{M})\right),
\end{equation}
respectively, where $\hat{O}_{\lambda \mu} (\mathcal{EM})$ is the operator in the long wave-length approximation with the rank $\lambda$ and its $z$-component $\mu$.
It is noted that the polarization for the magnetic transition is vertical to that for the electric one.

Using the matrix elements of Eq.~(\ref{eq:mele}), we obtain the electric dipole ($E1$) and magnetic dipole ($M1$) transition strengths of $^9$Be as
\begin{equation}
\frac{d^6B(EM1)_x}{d\mathbf{k}d\mathbf{K}} = \sum_{M_i} \frac{1}{2J_i + 1} \left| \mathcal{M}_x(EM1) \right|^2,
\end{equation}
where $J_i$ and $M_i$ are the total spin and its $z$-component of the initial ground state, respectively.
We obtain also the double-differential cross section for the $E1$ and $M1$ transitions of $^9$Be as
\begin{equation}
\begin{split}
\frac{d^2\sigma_x}{dE_\gamma d\Omega_n} =& \frac{16\pi^3}{9} \cdot \frac{E_\gamma}{\hbar c} \cdot \int d\mathbf{k} \int dK \frac{d^6B(EM1)_x}{d\mathbf{k}d\mathbf{K}}\\
&\times \delta\left(E_\gamma-E_\text{gs}-\frac{\hbar^2k^2}{2\mu}-\frac{\hbar^2K^2}{2M}\right),
\end{split}
\end{equation}
where $E_\gamma$ and $E_\text{gs}$ are the incident photon energy and the binding energy of the $^9$Be ground state, respectively.
The solid angle for the emitted neutrons is given by $\Omega_n$.
We here take the scattering angle $\theta_n$ of emitted neutrons as $\pi /2$.

When $\theta_n = \pi/2$, the cross section has a simple function form as a function of the azimuthal angle $\phi_n$ of the emitted neutrons~\cite{Agodi57}, that is,
\begin{equation}
\frac{d^2\sigma_x}{dE_\gamma d\Omega_n}\bigg|_{\theta = \frac{\pi}{2}} = a(E_\gamma)\left\{1+b(E_\gamma).\cos{2\phi_n}\right\}.
\end{equation}
It is noted that the coefficients $a$ and $b$ depend on $E_\gamma$.
In what follows we refer to the coefficient $b$ as the anisotropy parameter and discuss its dependence on the nuclear structure and transition modes in the next section.

\section{Results}
\begin{figure}[tb]
\centering{\includegraphics[clip,width=0.5\textwidth]{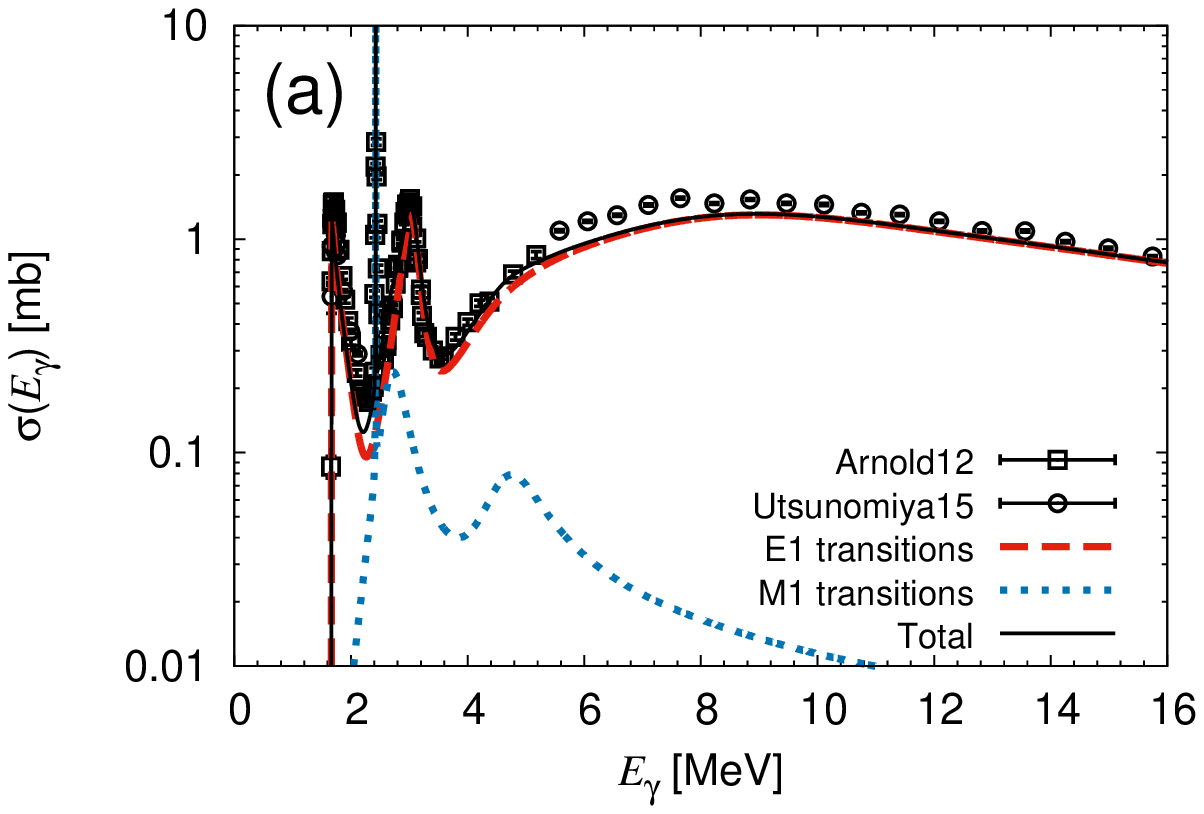}}
\centering{\includegraphics[clip,width=0.5\textwidth]{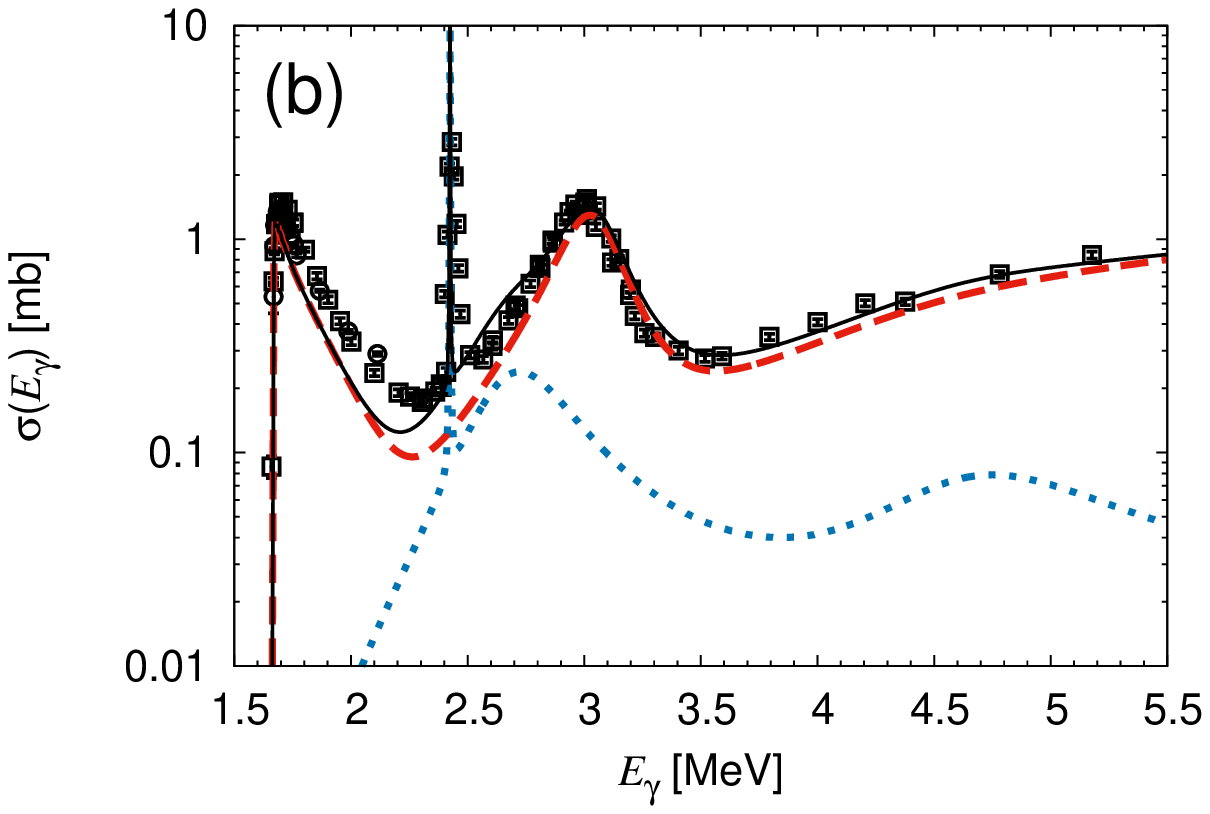}}
\caption{\label{fig:PDX}
(Color online) Calculated cross sections of the $^9$Be($\gamma$,$n$) reaction. The panel (a) represents the cross sections of up to $E_\gamma =$ 16 MeV and the panel (b) is the enlarged figure of the panel (a) in the range of $1.5 \le E_\gamma \le 5.5$ MeV.
The contributions of the $E1$ and $M1$ transitions are shown as the red (dashed) and blue (dotted) lines, respectively.
The sum of the contributions is shown as black (solid) line.
The open squares and open circles represent experimental data taken from Refs.~\cite{Arnold12} and \cite{Utsunomiya15}, respectively.
}
\end{figure}

First we show in Fig.~\ref{fig:PDX} the calculated cross section of the $^9$Be($\gamma$,$n$) reaction in comparison with the experimental data~\cite{Arnold12,Utsunomiya15}.
We show also the contributions of the $E1$ and $M1$ transitions.
From this comparison, it is seen that our calculation well reproduces the experimental data below $E_\gamma = 16$ MeV.
The calculated cross section below $E_\gamma=6$ MeV shows peaks coming from the resonances excited by the $E1$ and $M1$ transitions.
For reference, we list the excitation energies $E_x$ and the decay widths $\Gamma$ of the resonances obtained in the CSM in Table.~\ref{ta:res}.
The peaks at $E_\gamma = 2.4$ and $3.0$ MeV in Fig.~\ref{fig:PDX} correspond the $5/2^-$ and $5/2^+$ resonances obtained at 2.43 and 3.04 MeV, respectively. 
We find that the $1/2^-$ resonance at 2.68 MeV and the $3/2^-_2$ resonance at 4.65 MeV have the peaks in the contribution of the $M1$ transition but have minor contributions to the cross section.
The $3/2^+$ resonance at 4.69 MeV is not clearly identified both in the cross section and the contribution of the $E1$ transition because of its wide decay width of $\Gamma = 1.44$ MeV.
It is noted that the resonance pole of the first excited $1/2^+$ state is not obtained in the CSM, whereas our calculation reproduces the peak observed at $E_\gamma = 1.7$ MeV corresponding to the $1/2^+$ state (see Ref.~\cite{Odsuren15} for details).
Above $E_\gamma = 6$ MeV, the cross section is dominated by the $E1$ transition and the contributions of the $M1$ transition to the cross section is relatively small.
We confirm also that there is no resonance excited by the $E1$ or $M1$ transition above the $E_\gamma = 6$ MeV;
hence, the $E1$ transition into non-resonant continuum states dominates the cross section above $E_\gamma = 6$ MeV.
\begin{table}[tb]
\caption{\label{ta:res}
Excitation energies $E_x$ and decay widths $\Gamma$ for the resonances excited by the $E1$ and $M1$ transitions (units in MeV).
The observed data except for the $1/2^+$ state are taken from Ref.~\cite{Tilley04}. The date for the $1/2^+$ state are taken from Ref.~\cite{Arnold12}
}
\begin{tabular}{ccc}
\hline
$J^\pi$ & Present ( $E_x$, $\Gamma$ ) & Exp. ( $E_x$, $\Gamma$ ) \\
\hline
$1/2^+$ & - & ( 1.732$\pm$0.002, 0.213$\pm$0.006 ) \\
$5/2^-$ & ( 2.43, $\sim$3$\times$$10^{-4}$ ) & ( 2.4294, 7.8$\times$$10^{-4}$ ) \\
$1/2^-$ & ( 2.68, 0.495 ) & ( 2.78, 1.01 ) \\
$5/2^+$ & ( 3.04, 0.323 ) & ( 3.049, 0.282 ) \\
$3/2^+$ & ( 4.69, 1.44 ) & ( 4.704, 0.743 ) \\
$3/2^-_2$ & ( 4.65, 1.18 ) & ( 5.59, 1.33 ) \\
\hline
\end{tabular}
\end{table}

\begin{figure}[tb]
\centering{\includegraphics[clip,width=0.5\textwidth]{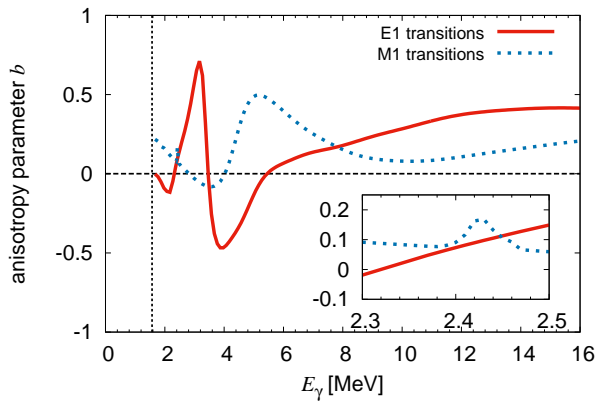}}
\caption{\label{fig:ani}
(Color online) Anisotropy parameters in the $^9$Be($\vec{\gamma}$,$n$) reaction.
Those for the $E1$ and $M1$ transitions are represented as the red (solid) and blue (dotted) lines, respectively.
The vertical line indicates the breakup threshold of $^9$Be.
The inset shows the enlarged figure for the energy region corresponding to the $5/2^{-}$ resonance.
}
\end{figure}
We calculate the azimuthal angle distribution of the neutrons emitted from the $^9$Be($\vec{\gamma}$,$n$) reaction and show the results in Fig.~\ref{fig:ani}.
In Fig.~\ref{fig:ani}, the anisotropy parameter for the $E1$ or $M1$ transition is plotted as a function of the photon incident energy $E_\gamma$.
In the low energy region below $E_\gamma = 6$ MeV, it is found that the anisotropy parameter fluctuates between positive and negative values as $E_\gamma$ varies.
The parameter for $E1$ has the maximal value at $E_\gamma = 3.2$ MeV corresponding to the $5/2^+$ resonance and that for $M1$ has the maximal value at $E_\gamma = 2.42$ MeV to the $5/2^-$ resonance.
Furthermore, it is seen that the anisotropy parameter for the $M1$ transition has the maximal value at the energy region of $E_\gamma \sim 5$ MeV.
In this energy region, the $M1$ transition is dominated by the transition into the $3/2^-_2$ resonance at 4.65 MeV with $\Gamma = 1.18$ MeV and the contribution of non-resonant continuum states is negligible.
Thus, the maximal value in the anisotropy parameter for the $M1$ transition at $E_\gamma \sim 5$ MeV is understood to come from the $3/2^-_2$ resonance. 
These results indicate that the anisotropy in the azimuthal angle distribution of the emitted neutrons is sensitive to the nuclear structure and is enhanced at the energies corresponding to the resonances in the final states.
It would be interesting that the anisotropy parameter for the $E1$ transition has a minimal value at $E_\gamma \sim 4$ MeV, which is coincide with the energy of the $3/2^+$ resonance at $E_x = 4.69$ MeV within the decay width of $\Gamma = 1.44$ MeV.
The $3/2^+$ resonance is not clearly identified even in the contribution of the $E1$ transition in Fig.~\ref{fig:PDX}.
This fact indicates the possibility that the azimuthal angle distribution of the neutrons emitted from the $^9$Be($\vec{\gamma}$,$n$) reaction is useful to identify the resonances not observed in the cross section of the ($\gamma$,$n$) reaction.
It is noted that the $1/2^\pm$ resonances cannot be clearly seen in the anisotropy parameter because the limitation on the $z$-component of the total spin suppresses the anisotropy in the azimuthal angle distribution.
In fact, the absolute values of the anisotropy parameters for the first excited $1/2^+$ state ($E_\gamma = 1.7$ MeV) in the $E1$ transition and the $1/2^-$ resonance ($E_\gamma = 2.68$ MeV) in the $M1$ transition are relatively small.

The transitions in the energy region higher than $E_\gamma = 6$ MeV are dominated by transitions into non-resonant continuum states, which have no peculiar structure.
Thus, it is expected that the anisotropy parameter is sensitive to the transition modes not to the nuclear structure.
As shown in Fig.~\ref{fig:ani}, the anisotropy parameter for the $E1$ or $M1$ transition gently changes.
The anisotropy for the $M1$ transition is lower than that for the $E1$ transition, but the anisotropy for the $M1$ transition has still positive value.
This result suggests that the sign of the anisotropy parameter does not depend on the transition modes in the non-resonant continuum region.

In the present work, we do not take into account the the effect of interferences between the $E1$ and $M1$ transitions on the anisotropy in the azimuthal angle distributions.
To discuss the azimuthal angle distributions more quantitatively, the analysis with the interference effect will be needed.

\section{Summary}

The electromagnetic transitions of $^9$Be with linearly polarized $\gamma$-rays were investigated by using the $\alpha$~+~$\alpha$~+~$n$ three-body model and the complex-scaled solutions of the Lippmann-Schwinger equation.
The anisotropy in the azimuthal angle distribution of the neutrons emitted from the $^9$Be($\vec{\gamma}$,$n$) reaction was calculated as a function of the photon incident energy $E_\gamma$.

We found that the anisotropy parameter for the $E1$ or $M1$ transition has maximal and minimal values in the low energy region below $E_\gamma = 6$ MeV, and their energies coincide with the resonance energies in the final states.
Although the $3/2^-_2$ and $3/2^+$ resonances have minor contributions to the photodisintegration cross section, these resonances have the maximal and minimal values, respectively, in the anisotropy parameter.
The azimuthal angle distributions of the neutrons emitted from the ($\vec{\gamma}$,$n$) reaction may be useful to identify the resonances not observed in the cross section of the ($\gamma$,$n$) reaction.
In contrast, in the energy region higher than $E_\gamma = 6$ MeV, we found that the anisotropy parameter gently changes and the signs for the $E1$ and $M1$ transitions are identical.
This result suggested that the anisotropy does not depend on the transition modes.

\section*{Acknowledgements}
This work was supported by JSPS KAKENHI Grant Numbers JP15H03665.

\bibliographystyle{apsrev}
\bibliography{References}

\end{document}